\documentclass[aps,prl,twocolumn]{revtex4}

\usepackage{graphicx,amsmath}

\preprint{v04}

\begin{document}

\title{Giant gyrotropy due to electromagnetic coupling}

\author{A. V. Rogacheva}
\affiliation{EPSRC NanoPhotonics Portfolio Centre, School of Physics and Astronomy, University of Southampton, SO17 1BJ, UK}

\author{V. A. Fedotov}
\email{vaf@phys.soton.ac.uk}
\affiliation{EPSRC NanoPhotonics Portfolio Centre, School of Physics and Astronomy, University of Southampton, SO17 1BJ, UK}

\author{A. S. Schwanecke}
\affiliation{EPSRC NanoPhotonics Portfolio Centre, School of Physics and Astronomy, University of Southampton, SO17 1BJ, UK}

\author{N. I. Zheludev}
\homepage{www.nanophotonics.org.uk}
\affiliation{EPSRC NanoPhotonics Portfolio Centre, School of Physics and Astronomy, University of Southampton, SO17 1BJ, UK}

\date{\today}

\pacs{n/a}

\begin{abstract}
We report the first experimental evidence that electromagnetic
coupling between physically separated planar metal patterns
located in parallel planes provides for exceptionally strong
polarization rotatory power if one pattern is twisted in respect
to the other, creating a 3D chiral object. In terms of optical
rotary power per sample of thickness equal to one wavelength, the
bi-layered structure rotates five orders of magnitude stronger
than a gyrotropic crystal of quartz in the visible spectrum. We
also saw a signature of negative refraction for circularly
polarized waves propagating through the chiral slab.
\end{abstract}

\maketitle

\begin{figure}
\includegraphics[width=3in]{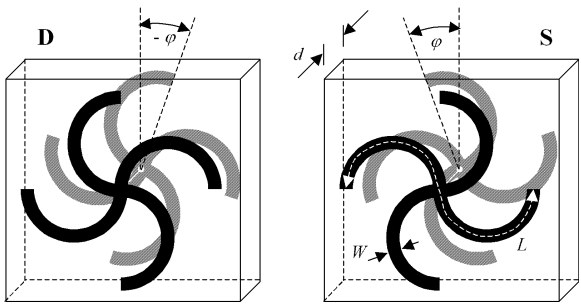}
\caption{Dextral (\textbf{D}, right-handed) and sinistral (\textbf{S}, left-handed) enantiomeric helicoidal bi-layered structures constructed from planar metal rosettes separated by a dielectric slab of thickness $d$. $L$ is the end-to-end length of the rosette's  strip and $W$ is its width. The rosettes forming the enantiomeric structures are mutually twisted around the axis of 4-fold symmetry on angles $-\varphi$ and $\varphi$ respectively.}
\end{figure}

\begin{figure}
\includegraphics[width=3.375in]{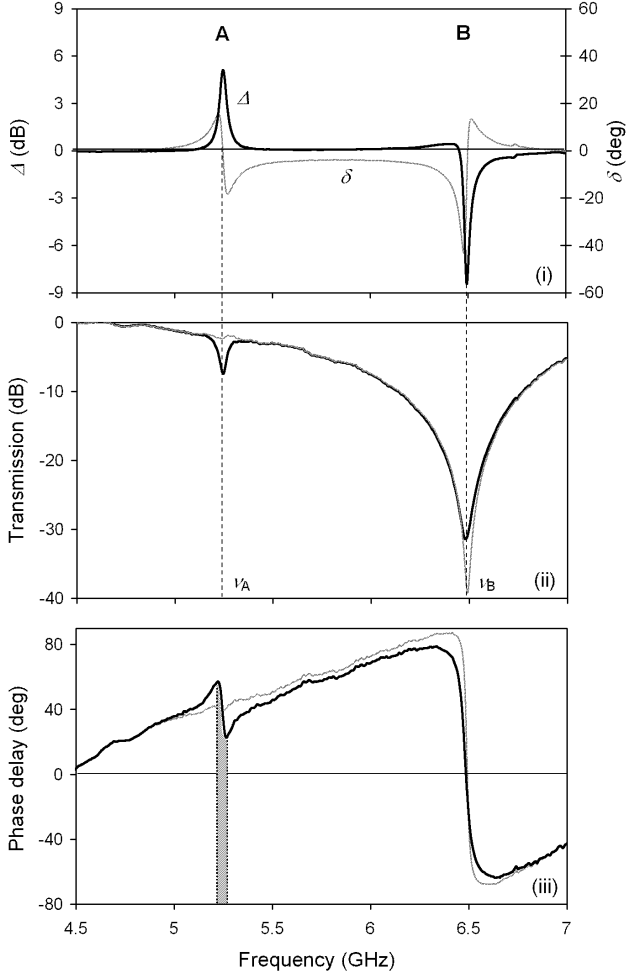}
\caption{Electromagnetic properties of the bi-layered sinistral
chiral structure with mutual twist $\varphi = 15^\circ$. Frequency dependencies of: (i) circular dichroism, $\Delta$, (black line) and circular differential phase delay, $\delta$, (gray line); (ii) transmission losses for right circularly polarized wave (black line) and left circularly polarized wave (gray line); (iii) phase delay for right circularly polarized wave (black line) and left circularly polarized wave (gray line). The shaded area in section (iii) represents a frequency range where the phase velocity $v_\mathrm{p}$ and group velocity $v_\mathrm{g}$ have opposite signs for right circular polarization, which is a signature of negative refraction.}
\end{figure}

The ability to rotate the polarization state of light (gyrotropy)
by chiral molecules is one of the most fundamental phenomena of
electrodynamics. It was discovered by F.~Aragot in 1811 and is now
widely used in analytical chemistry, biology and crystallography
for identifying the spatial structure of molecules. Recent explosive
increase in the interest in gyrotropic media is driven by an
opportunity for the development of negative index meta-materials,
where simultaneous electric and magnetic response of gyrotropic
structures are required to achieve negative refraction
\cite{Tretyakov1,Pendry,Tretyakov2,Jin,Monzon,Agranovich,Cheng}.
Sculptured helical pillars for the optical part of the spectrum
\cite{Pillars}, helical wire springs \cite{Springs1,Springs2} and
twisted Swiss-role metal structures \cite{Pendry} for microwave applications have been discussed as possible candidates
for achieving strong artificial gyrotropy that can be used for
implementing negative refraction. However, from meta-material prospective it would be very desirable if
chirality could be achieved by planar patterning using
well-established planar technologies, thus making nano-fabrication
of such structures for the optical part of the spectrum a
practical proposition. The opportunity of creating true 3D
chirality in non-contacting layers of planar metal structures was
first identified in Ref.~\cite{Twisted_Bars}. It was suggested
that inductive coupling between two identical mutually twisted
metal patterns can create an optically active chiral object and
thus provide for gyrotropy.

In this letter we show the first experiential demonstration that
giant optical gyrotropy can be achieved in a bi-layered chiral
structure through electromagnetic coupling between the layers and
that there is no need to sculpture continuous helix-like volume
three-dimensional chiral objects to achieve strong polarization
rotatory power. We also saw clear evidence of negative
refraction in the structure. The experiments were performed in the
microwave part of the spectrum. Although we expect the effect to
be seen with a large variety  of patterns, we investigated a
structure consisting of two identical metal rosettes of 4-fold
rotational symmetry located in parallel planes, as presented on Fig.~1. The 4-fold symmetry of the rosette ensures that the structure is isotropic for observations at normal incidence and therefore shows no birefringence. Due to curved lines rosette-like structure exhibit resonant properties at wavelengths larger than the overall size of the design. The latter would be important for achieving a non-diffracting regime if two-dimensional arrays of such
structures were used to form planar metamaterial sheets and volume
structures. The rosettes were etched from $35~\mu\text{m}$ flat
copper film on both sides of a dielectric substrate. The rosettes
had the length $L = 53~\text{mm}$ and strip width $W =
0.4~\text{mm}$ and were spaced by a homogeneous dielectric layer
of thickness $d = 1.5~\text{mm}$ ($\varepsilon= 3.77+i\,0.03$).

We studied circular birefringence and dichroism of the structure
in $4.5 - 7.0~\text{GHz}$ frequency range (wavelength range $4.3 -
6.7~\text{cm}$) using a microwave waveguide polarimeter. The
polarimeter included a $480~\text{mm}$ long circular waveguide
with a diameter of $41.5~\text{mm}$ and two high quality circular
polarizers of either same or opposite handedness (series 64 by Flann
Microwave) attached on both sides of the waveguide. 
Each sample
was placed in the middle of the waveguide, perpendicular to its
axis. A full S-parameter vector network analyzer (model E8364B by
Agilent) was used to measure both magnitude and phase of the
wave transmitted through the polarimeter.

In our experiments we compared pairs of enantiomeric forms of the
structure (designated as type \textbf{D} and \textbf{S} in Fig.~1)
with various angles of mutual twist, $\varphi$, in the range from
$\varphi=0^\circ$ to $\varphi=45^\circ$. To describe the results
of polarimetric measurements we will define transmission of a
sample, $t$, measured by the polarimeter as follows: the
superscript index refers to the type of the sample, \textbf{D} or
\textbf{S}, while subscript indices refer to the state of
polarizer and analyzer. In these terms circular dichroism is
defined as $\Delta$ = $|t^{\mathrm{S}}_{++}|^2 -
|t^{\mathrm{S}}_{--}|^2$ while circular differential phase delay
(responsible for circular birefringence) is defined as $\delta$ =
$\arg(t^{\mathrm{S}}_{++}) - \arg(t^{\mathrm{S}}_{--})$. To
eliminate any possible polarization effects, which could have
resulted not from three-dimensional chirality, but from anisotropic imperfections of polarimeter and/or sample, we
performed experiments at different mutual orientations rotating the
sample around the axis of the cylindrical waveguide,
and found virtually no dependence of the observed effects on the
orientation.

The structure's three-dimensional chirality and thus its
gyrotropic characteristics should depend strongly on the mutual
orientation of the rosettes and distance $d$ between them. To
verify this we manufactured a truly planar version of the
structures by etching both rosettes from the same metal film
($d=0$) and sandwiching them between two layers of dielectric for
symmetry. We also manufactured a bi-layered structure with no twist
between rosettes ($\varphi =0$, $d \neq 0$). No circular dichroism or
differential phase delay was observed in both cases. We also
studied the dependence of gyrotropic characteristics of the
structure on the value of $\varphi$.  The complete picture of the
dependence of gyrotropy on $\varphi$ may be obtained by measuring
it in the interval of $\varphi$ between $0$ and $45^\circ$.
Indeed, due to the four-fold symmetry of the rosettes, the structure
with $\varphi$ between $0$ and $45^\circ$ is equivalent to the
structures twisted on $\varphi \pm n \cdot 90^\circ$, where $n$ is
an integer number. The sinistral structure with $\varphi$
in between $45^\circ$ and $90^\circ$ is equivalent to 
the dextral structure with mutual twist of $\varphi - 90^\circ$ and therefore has an opposite sign of gyrotropy.

Characteristic spectral dependencies of $\Delta$ and $\delta$
exhibited by the sinistral structure are presented in Fig.~2(i)
for $\varphi =15^\circ$. They show two resonances located below
and above $6~\text{GHz}$. They will be called resonance $A$ and
$B$ correspondingly. 
From Fig.~2(ii) it follows that on a
sinistral (left) structure, losses at resonance $A$ for right
circular polarization are smaller than for left circular
polarization. However, losses at resonance $B$ for right circular
polarization are stronger than for left circular polarization.
Phase delay $\psi= \arg(t)$ is presented on Fig.~2(iii). Provided that an electromagnetic wave is presented in the form $e^{i(\omega t + kx)}$, in the proximity of resonance $A$ the group velocity $v_\mathrm{g} = d/(\partial \psi / \partial\omega)$ for right circular polarization in the sinistral structure has opposite sign to the phase velocity $v_\mathrm{p} = c \psi \lambda_g / 2 \pi d$ (here $\lambda_g$ is guided the wavelength in the circular waveguide). In accordance with Ref. \cite{Pendry} this is a signature
of negative refraction in chiral media. In proximity of resonance $B$, negative refraction is observed for both circular polarizations on the background of high losses. Availability of negative refraction is especially important in the proximity of resonance $A$, where overall losses are small.

\begin{figure*}
\includegraphics[width=6.75in]{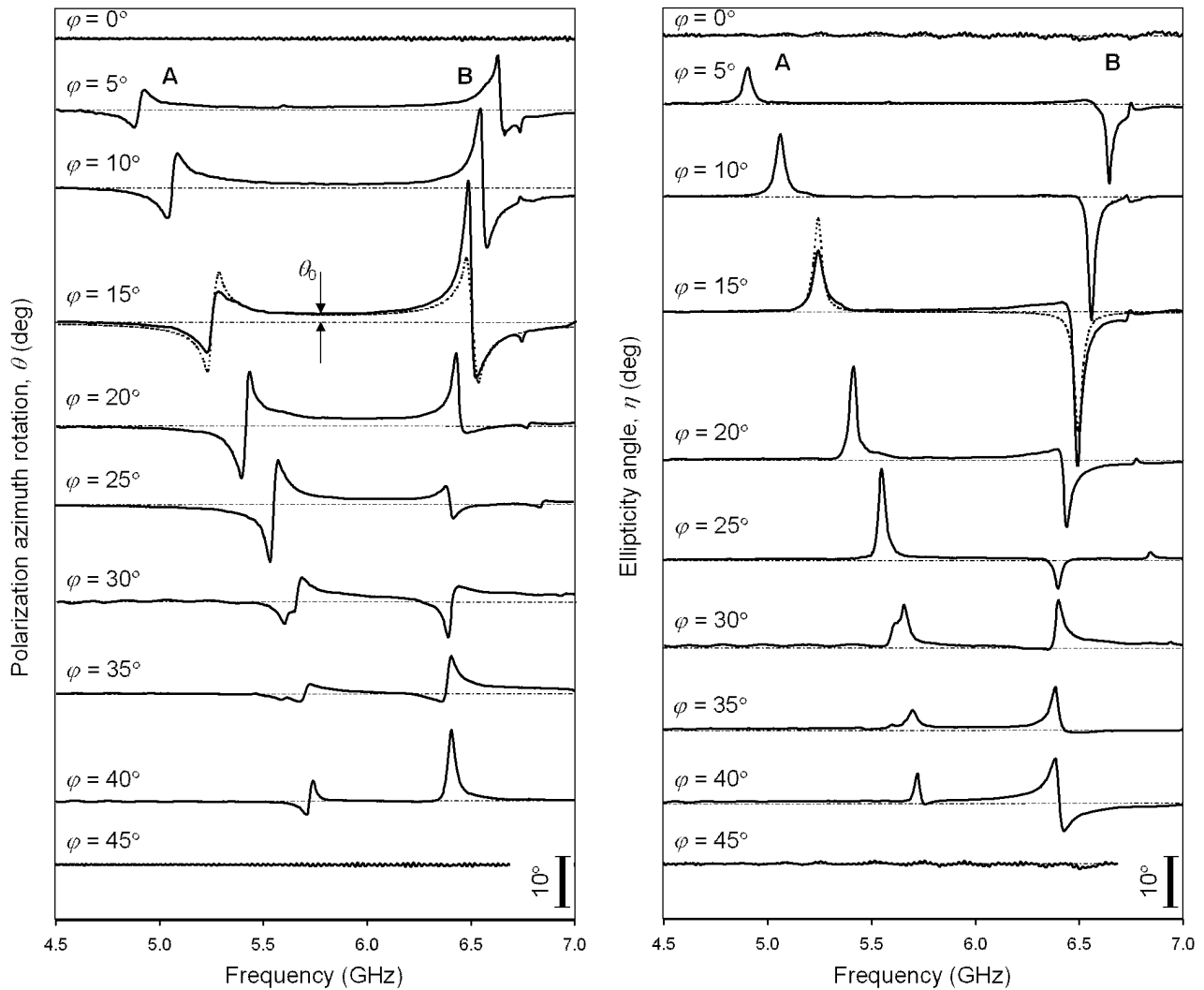}
\caption{Frequency dependencies of polarization azimuth rotation
$\theta$ (left column) and ellipticity, in terms of the
ellipticity angle $\eta$ (right column) that a linearly polarized
wave would acquire upon transmission through a sinistral bi-layered chiral structure for various twist angles $\varphi$. Vertical
scale is shown in the right bottom corners. Dotted line at
$\varphi=15^\circ$ shows the frequency dispersion of the effect as
predicted by the Born-Kuhn model.}
\end{figure*}


In general, a linearly polarized wave transmitted through the
structure will become elliptical on transmission and its
polarization azimuth will rotate. Using standard definitions of
the degree for ellipticity $\eta$ and polarization azimuth rotation
$\theta$ of elliptically polarized light \cite{Jackson}, we
calculated polarization changes of the linearly polarized incident
wave as follows: $\eta =  \frac{1}{2}
\arcsin(\frac{\Delta}{|t^{\mathrm{S}}_{++}|^2 +
|t^{\mathrm{D}}_{++}|^2})$,~ $\theta = - \frac{1}{2}\delta$. This
is presented in Fig.~3 for different values of
$\varphi$. The peak values of both rotation and ellipticity
initially increase with $\varphi$. They reach their absolute
maxima at about $\varphi =15^\circ$ with $\eta =-30^\circ$ at
frequency $\nu_B$ ($\eta = -45^\circ$ corresponds to perfectly
left-handed circularly polarized light). At the exact resonance no
polarization rotation is seen as its dispersion passes zero, but in
proximity of resonance $\nu_B$ rotation reaches
$\theta_{\mathrm{B}^{+}}=28^\circ$.
Between the peaks, in the
spectral range of low losses and virtually zero dichroism, we
observe a pure rotation of polarization azimuth of about $\theta_0
=3^\circ$. With further increase of $\varphi$, resonances A and B
move closer to one another and the dispersions of $\theta$ and $\eta$
change. Peak value of rotation and ellipticity decreases as well as
rotatory power at frequencies between the peaks. Gyrotropy
completely collapses at $\varphi =45^\circ$ as should be expected, see Ref.~\cite{Twisted_Bars}.

It shall be noted that the observed rotation induced by the
artificial bi-layered structure (which has a thickness $d$ of only
about 1/30 of the wavelength) is huge. To appreciate its magnitude
it shall be compared with the gyrotropy of natural optical active
materials. Indeed, in terms of optical rotary power per sample
thickness equal to one wavelength, bi-layered structure rotates
\emph{five orders of magnitude stronger} than a gyrotropic crystal
of quartz and \emph{three orders of magnitude stronger} than cholesteric liquid crystals in the visible spectrum (specific rotatory power of quartz and cholesteric liquid crystals is about $20^\circ/\text{mm}$ and $10^{3~\circ} /\text{mm}$ \cite{Cholesterics} respectively). Rotatory power of the bi-layered system is also two orders of magnitude stronger that in the recently introduced metal-on-dielectric chiral system, where resonant rotation of about $1^\circ$ was seen in a sample $1/6$ of the wavelength thick \cite{Gonokami}.

To identify the underlying nature of the observed polarization
effect we recall the classical model of gyrotropy developed
by Born and Kuhn \cite{Book}. In this model two spatially separated charged oscillators moving along orthogonal directions have an elastic binding between them. Excitation of one of them by the incident electromagnetic wave is then transferred by the elastic coupling to the other. Induced oscillations of its charge then re-emit a wave at a different polarization and with some delay,
thus ensuring both polarization azimuth rotation and dichroism.
Analogously, current driven in the rosette arm by the incident
wave is inductively (or capacitively) coupled to the current in
the rosette arm of the second layer. The induced current in the
second layer is then re-emited into the transmitted wave with a
different polarization state providing for gyrotropy. It is
remarkable how well the Born-Kuhn model is suitable for describing
this process. It gives the following dispersion of the effect
$\binom{\theta}{\eta} \propto \binom{\text{Re}}{\text{Im}} \,
\frac{\xi \nu^2}{(\nu_0^2+\xi - i \gamma \nu - \nu^2)^2-\xi^2}$,
where $\nu_0$ and $\gamma$ are the resonant frequency and damping
parameter of individual oscillator and $\xi$ is the Hooke
coefficient of elastic force between the oscillators \cite{Book}.
Here, $\nu_0$ is analogous to the resonant frequency of the dipole
interaction of an electromagnetic wave with the arm of an individual rosette, which shall be about $c/L = 5.7~\text{GHz}$. The Hooke elastic interaction between the oscillators in the Borh-Kuhn model is analogous to the electromagnetic coupling between rosettes.
According to above formula, the Born-Kuhn model predicts a two-peak dispersion of rotatory power and circular dichroism with the
peak spectral separation increasing with coupling $\xi$. Indeed,
in our experiments the peak separation is at maximum for small
$\varphi$ when the electromagnetic interaction between rosettes is
strong. It decreases with increasing $\varphi$ when rosette
overlapping diminishes and separation of the peaks reduces
accordingly. The Born-Kuhn dispersion accurately describes the main
features of the rotatory power and circular dichroism in the
bi-layered structure as may be see in Fig.3, where
theoretical dispersion curves are plotted as a dotted line for
$\varphi =15^\circ$ ($\nu_0=5.247~\text{GHz}$, $\gamma=4.4 \cdot
10^7 \text{s}^{-1}$, $\xi = 4.6 \cdot 10^{19} \text{s}^{-2}$).
Importantly, the model indicates strong coupling between rosettes
in the bi-layered system, where figure of merit is $\xi/\nu_0^2 =
0.26$, i.e. the energy of the interaction between
rosettes amounts to a quarter of the energy of interaction between
individual rosette and field, which explains incredibly strong
gyrotropy of the system in chiral configurations. The Born-Kuhn
model is less accurate in giving a correct ratio of the effect
magnitude in the peaks. This small discrepancy is not surprising
as elastic coupling is not really equivalent to the
electromagnetic one, and there are other mechanism of enantiomericly
sensitive interactions with the bi-layered structure, which are not
covered by the Born-Kuhn model. For instance, one can see the
twisted rosette pair as an enantiomeric sensitive scattering
object. Scattering, could happen in all direction creating
enantiomericly sensitive losses for the wave propagating in the
forward direction. In the reality of the confined environment of the
waveguide we perhaps see a strong interplay between the Born-Kuhn
like gyrotropy and enantiomericly sensitive scattering that creates a
complex frequency dispersion of the effect at various $\varphi$.

In summary, we provided the first experimental evidence that
strong gyrotropy can be achieved by electromagnetic coupling in
chiral bi-layered disconnected metal structures and saw a signature
of negative refraction for circularly polarized waves on the chiral
bi-layered structure. We expect that optical activity of this
nature will also be displayed by appropriately scaled
sub-wavelength nanostructures in the optical part of the spectrum.

\begin{acknowledgments}
The authors are grateful to Martin McCall for fruitful discussions. 
Financial support of the Engineering and Physical Sciences
Research Council, UK and EU Network of Excellence "Metamorphose"
is acknowledged.
\end{acknowledgments}

\bibliography{Giant_Gyrot_EMC}

\begin{thebibliography}{15}
\expandafter\ifx\csname natexlab\endcsname\relax\def\natexlab#1{#1}\fi
\expandafter\ifx\csname bibnamefont\endcsname\relax
  \def\bibnamefont#1{#1}\fi
\expandafter\ifx\csname bibfnamefont\endcsname\relax
  \def\bibfnamefont#1{#1}\fi
\expandafter\ifx\csname citenamefont\endcsname\relax
  \def\citenamefont#1{#1}\fi
\expandafter\ifx\csname url\endcsname\relax
  \def\url#1{\texttt{#1}}\fi
\expandafter\ifx\csname urlprefix\endcsname\relax\def\urlprefix{URL }\fi
\providecommand{\bibinfo}[2]{#2}
\providecommand{\eprint}[2][]{\url{#2}}

\bibitem[{\citenamefont{Tretyakov et~al.}(2003)\citenamefont{Tretyakov,
  Nefedov, Sihvola, Maslovski, and Simovski}}]{Tretyakov1}
\bibinfo{author}{\bibfnamefont{S.}~\bibnamefont{Tretyakov}},
  \bibinfo{author}{\bibfnamefont{I.}~\bibnamefont{Nefedov}},
  \bibinfo{author}{\bibfnamefont{A.}~\bibnamefont{Sihvola}},
  \bibinfo{author}{\bibfnamefont{S.}~\bibnamefont{Maslovski}},
  \bibnamefont{and} \bibinfo{author}{\bibfnamefont{C.}~\bibnamefont{Simovski}},
  \bibinfo{journal}{J. Electromagn. Waves Appl.} \textbf{\bibinfo{volume}{17}},
  \bibinfo{pages}{695} (\bibinfo{year}{2003}).

\bibitem[{\citenamefont{Pendry}(2004)}]{Pendry}
\bibinfo{author}{\bibfnamefont{J.~B.} \bibnamefont{Pendry}},
  \bibinfo{journal}{Science} \textbf{\bibinfo{volume}{306}},
  \bibinfo{pages}{1353} (\bibinfo{year}{2004}).

\bibitem[{\citenamefont{Tretyakov et~al.}(2005)\citenamefont{Tretyakov,
  Sihvola, and Jylha}}]{Tretyakov2}
\bibinfo{author}{\bibfnamefont{S.}~\bibnamefont{Tretyakov}},
  \bibinfo{author}{\bibfnamefont{A.}~\bibnamefont{Sihvola}}, \bibnamefont{and}
  \bibinfo{author}{\bibfnamefont{L.}~\bibnamefont{Jylha}},
  \bibinfo{journal}{Photonics Nanostruct. Fundam. Appl.}
  \textbf{\bibinfo{volume}{3}}, \bibinfo{pages}{107} (\bibinfo{year}{2005}).

\bibitem[{\citenamefont{Jin and He}(2005)}]{Jin}
\bibinfo{author}{\bibfnamefont{Y.}~\bibnamefont{Jin}} \bibnamefont{and}
  \bibinfo{author}{\bibfnamefont{S.}~\bibnamefont{He}}, \bibinfo{journal}{Opt.
  Express} \textbf{\bibinfo{volume}{13}}, \bibinfo{pages}{4974}
  (\bibinfo{year}{2005}).

\bibitem[{\citenamefont{Monzon and Forester}(2005)}]{Monzon}
\bibinfo{author}{\bibfnamefont{C.}~\bibnamefont{Monzon}} \bibnamefont{and}
  \bibinfo{author}{\bibfnamefont{D.~W.} \bibnamefont{Forester}},
  \bibinfo{journal}{Phys. Rev. Lett.} \textbf{\bibinfo{volume}{95}},
  \bibinfo{pages}{123904} (\bibinfo{year}{2005}).

\bibitem[{\citenamefont{Agranovich et~al.}(2006)\citenamefont{Agranovich,
  Gartstein, and Zakhidov}}]{Agranovich}
\bibinfo{author}{\bibfnamefont{V.~M.} \bibnamefont{Agranovich}},
  \bibinfo{author}{\bibfnamefont{Y.~N.} \bibnamefont{Gartstein}},
  \bibnamefont{and} \bibinfo{author}{\bibfnamefont{A.~A.}
  \bibnamefont{Zakhidov}}, \bibinfo{journal}{Phys. Rev. B}
  \textbf{\bibinfo{volume}{73}}, \bibinfo{pages}{045114}
  (\bibinfo{year}{2006}).

\bibitem[{\citenamefont{Cheng and Cui}(2006)}]{Cheng}
\bibinfo{author}{\bibfnamefont{Q.}~\bibnamefont{Cheng}} \bibnamefont{and}
  \bibinfo{author}{\bibfnamefont{T.~J.} \bibnamefont{Cui}},
  \bibinfo{journal}{Phys. Rev. B} \textbf{\bibinfo{volume}{73}},
  \bibinfo{pages}{113104} (\bibinfo{year}{2006}).

\bibitem[{\citenamefont{Lakhtakia and Messier}(2005)}]{Pillars}
\bibinfo{author}{\bibfnamefont{A.}~\bibnamefont{Lakhtakia}} \bibnamefont{and}
  \bibinfo{author}{\bibfnamefont{R.}~\bibnamefont{Messier}},
  \emph{\bibinfo{title}{Sculptured Thin Films: Nanoengineered Morphology and
  Optics}} (\bibinfo{publisher}{SPIE Press, Bellingham, WA},
  \bibinfo{year}{2005}).

\bibitem[{\citenamefont{Tinoco and Freeman}(1957)}]{Springs1}
\bibinfo{author}{\bibfnamefont{I.}~\bibnamefont{Tinoco}} \bibnamefont{and}
  \bibinfo{author}{\bibfnamefont{M.~P.} \bibnamefont{Freeman}},
  \bibinfo{journal}{J. Phys. Chem.} \textbf{\bibinfo{volume}{61}},
  \bibinfo{pages}{1196} (\bibinfo{year}{1957}).

\bibitem[{\citenamefont{Jaggard and Engheta}(1989)}]{Springs2}
\bibinfo{author}{\bibfnamefont{D.~L.} \bibnamefont{Jaggard}} \bibnamefont{and}
  \bibinfo{author}{\bibfnamefont{N.}~\bibnamefont{Engheta}},
  \bibinfo{journal}{Electron. Lett.} \textbf{\bibinfo{volume}{25}},
  \bibinfo{pages}{173} (\bibinfo{year}{1989}).

\bibitem[{\citenamefont{Svirko et~al.}(2001)\citenamefont{Svirko, Zheludev, and
  Osipov}}]{Twisted_Bars}
\bibinfo{author}{\bibfnamefont{Y.}~\bibnamefont{Svirko}},
  \bibinfo{author}{\bibfnamefont{N.}~\bibnamefont{Zheludev}}, \bibnamefont{and}
  \bibinfo{author}{\bibfnamefont{M.}~\bibnamefont{Osipov}},
  \bibinfo{journal}{Appl. Phys. Lett.} \textbf{\bibinfo{volume}{78}},
  \bibinfo{pages}{498} (\bibinfo{year}{2001}).

\bibitem[{\citenamefont{Jackson}(1999)}]{Jackson}
\bibinfo{author}{\bibfnamefont{J.~D.} \bibnamefont{Jackson}},
  \emph{\bibinfo{title}{Classical Electrodynamics}} (\bibinfo{publisher}{John
  Wiley, New York}, \bibinfo{year}{1999}), \bibinfo{edition}{3rd} ed.

\bibitem[{\citenamefont{de~Gennes}(1974)}]{Cholesterics}
\bibinfo{author}{\bibfnamefont{P.~G.} \bibnamefont{de~Gennes}},
  \emph{\bibinfo{title}{The physics of liquid crystals}}
  (\bibinfo{publisher}{Clarendon Press, Oxford}, \bibinfo{year}{1974}).

\bibitem[{\citenamefont{Kuwata-Gonokami
  et~al.}(2005)\citenamefont{Kuwata-Gonokami, Saito, Ino, Kauranen, Jefimovs,
  Vallius, Turunen, and Svirko}}]{Gonokami}
\bibinfo{author}{\bibfnamefont{M.}~\bibnamefont{Kuwata-Gonokami}},
  \bibinfo{author}{\bibfnamefont{N.}~\bibnamefont{Saito}},
  \bibinfo{author}{\bibfnamefont{Y.}~\bibnamefont{Ino}},
  \bibinfo{author}{\bibfnamefont{M.}~\bibnamefont{Kauranen}},
  \bibinfo{author}{\bibfnamefont{K.}~\bibnamefont{Jefimovs}},
  \bibinfo{author}{\bibfnamefont{T.}~\bibnamefont{Vallius}},
  \bibinfo{author}{\bibfnamefont{J.}~\bibnamefont{Turunen}}, \bibnamefont{and}
  \bibinfo{author}{\bibfnamefont{Y.}~\bibnamefont{Svirko}},
  \bibinfo{journal}{Phys. Rev. Lett.} \textbf{\bibinfo{volume}{95}},
  \bibinfo{pages}{227401} (\bibinfo{year}{2005}).

\bibitem[{\citenamefont{Svirko and Zheludev}(1998)}]{Book}
\bibinfo{author}{\bibfnamefont{Y.~P.} \bibnamefont{Svirko}} \bibnamefont{and}
  \bibinfo{author}{\bibfnamefont{N.~I.} \bibnamefont{Zheludev}},
  \emph{\bibinfo{title}{Polarization of Light in Nonlinear Optics}}
  (\bibinfo{publisher}{John Wiley, Chichester}, \bibinfo{year}{1998}).

\end{thebibliography}

\end{document}